\documentclass{ragtime}

\title[Extended resonance model]%
      {Extended orbital resonance model with hump-induced oscillations}

\author[Z. Stuchl\'{\i}k, P. Slan\'y and G. T\"or\"ok]%
      {Zden\v{e}k Stuchl\'{\i}k\at{a} Petr Slan\'y and Gabriel T\"or\"ok\\%
      \iophaffil[]\\%
      \ins{a}\Email{zdenek.stuchlik@fpf.slu.cz}}

\coentry{Z. Stuchl\'{\i}k, P. Slan\'y and G. T\"or\"ok}
      {Extended orbital resonance model with hump-induced oscillations}

\begin{document}

\vspace{-1.0\baselineskip}

\begin{abstract}
  Change of sign of the LNRF-velocity gradient has been found for accretion
  discs orbiting rapidly rotating Kerr black holes with spin $a>0.9953$ for
  Keplerian discs and $a>0.99979$ for marginally stable thick discs. Such a
  ``humpy'' LNRF-velocity profiles occur just above the marginally stable
  circular geodesic and could be related to oscillations of accretion discs.
  The frequency of such ``hump''-induced oscillations can be identified with
  the maximal rate of change of the orbital velocity within the ``humpy''
  profile.  Therefore, we introduce an extended orbital resonance model
  (EXORM) of quasiperiodic oscillations (QPOs) assuming non-linear resonant
  phenomena between oscillations with the orbital epicyclic frequencies and
  the humpy frequency defined in a fully general relativistic way. The EXORM
  is developed for both Keplerian discs and perfect-fluid tori where the
  approximation of oscillations with epicyclic frequencies is acceptable.
  Clearly, the EXORM could be applied to the near-extreme Kerr black hole
  systems exhibiting relatively complex QPO frequency patterns. Assuming a
  Keplerian disc, it can be shown that in the framework of the EXORM, all the
  QPOs observed in the microquasar GRS 1915$+$105 could be explained, while it
  is not possible in the case of QPOs observed in the Galactic Centre source
  Sgr\,A$^*$.
\end{abstract}


\vspace{-1.0\baselineskip}

\section{Introduction}\label{intro}

High frequency (kHz) twin peak quasi-periodic oscillations (QPOs) with
frequency ratios $3\!:\!2$ (and sometimes 3:1) are observed in
microquasars~\citep[see,
e.g.,][]{Kli:2000:ARASTRA:,McCli-Rem:2004:CompactX-Sources:,Rem:2005:ASTRN:}.
In the Galactic Centre black hole Sgr\,A$^*$, \citet{Gen-etal:2003:NATURE:}
measured a clear periodicity of 1020~sec in variability during a flaring
event. This period is in the range of Keplerian orbital periods at a few
gravitational radii from a black hole with mass $M\sim 3.6\times
10^{6}\,\msun$ estimated for
Sgr\,A$^*$~\citep{Ghe-etal:2005:ASTRJ2:,Wei-Mil-Ghe:2005:ASP338:astro-ph0512621}.
More recently, \citet{Asc-etal:2004:ASTRA:,Asch:2004:ASTRA:,Asc:2006:CHIAA:}
reported three QPO periodicities at $692\,\mathrm{sec}$, $1130\,\mathrm{sec}$
and $2178\,\mathrm{sec}$ that correspond to frequency ratios
$(1/692)\!:\!(1/1130)\!:\!(1/2178)\sim 3\!:\!2\!:\!1$.  These observational
data are not quite convincing~\citep[see,
e.g.,][]{Abr-etal:2004:RAGtime4and5:CrossRef}, but surely deserve
attention~\citep{Asch:2007:CHIAA:MassSpin}.

{\sloppy
  Detailed analysis of the variable X-ray black-hole binary system
  (mi\-cro\-quas\-ar) GRS 1915$+$105 reveals high-frequency QPOs appearing at
  five frequencies, namely \mbox{$\nu_{1}=(27\pm
    1)\,\mathrm{Hz}$}~\citep{Bel-Men-San:2001:ASTRA:HFQPOs},
  \mbox{$\nu_{2}=(41\pm 1)\,\mathrm{Hz}$}, \mbox{$\nu_{3}=(67\pm
    1)\,\mathrm{Hz}$}~\citep{Mor-Rem-Gre:1997:ASTRJ2:,Str:2001:ASTRJ2:554},
  and \mbox{$\nu_{4}=(113\pm 5)\,\mathrm{Hz}$}, \mbox{$\nu_{5}=(167\pm
    5)\,\mathrm{Hz}$}~\citep{Rem-McCli:2006:ARASTRA:}. In this range of their
  errors, both upper pairs are close to the frequency ratio $3\!:\!2$
  suggesting the possible existence of resonant phenomena in the system.
  Observations of oscillations with these frequencies have different
  qualities, but in all five cases the data are quite
  convincing~\mbox{\citep[see][]{McCli-Rem:2004:CompactX-Sources:,Rem-McCli:2006:ARASTRA:}}.
\par}

Several models have been developed to explain the kHz QPO frequencies, and it
is usually preferred that these oscillations are related to the orbital motion
near the inner edge of an accretion disc. In particular, two ideas based on
the strong-gravity properties have been proposed.
While~\citet{Ste-Vie:1998:ASTRJ2L:,Ste-Vie:1999:PHYRL:} introduced the
``Relativistic Precession Model'' considering that the kHz QPOs directly
manifest the modes of a slightly perturbed (and therefore epicyclic)
relativistic motion of blobs in the inner parts of the accretion disc,
\citet{Klu-Abr:2001:ACTPB:} propose models based on non-linear oscillations of
an accretion disc that assume resonant interaction between orbital and/or
epicyclic modes. In a different context, the possibility of resonant coupling
between the epicyclic modes of motion in the Kerr spacetime was also mentioned
in the early work of~\citet{Ali-Gal:1981:GENRG2:}.  The radial and vertical
epicyclic oscillations could be related to both the thin Keplerian
discs~\citep{Abr-etal:2003:PUBASJ:,Kat:2001:PUBASJ:b} and the thick, toroidal
accretion discs~\citep{Rez-etal:2003:MONNR:}.  In particular, the observations
of high frequency twin peak QPOs with the $3\!:\!2$ frequency ratio in
microquasars can be explained by the parametric resonance between the radial
and vertical epicyclic oscillations,
$\nu_{\mathrm{v}}\!:\!\nu_{\mathrm{r}}\sim 3\!:\!2$.  This hypothesis, under
the assumption of geodesic oscillations (i.e., for thin discs), puts strong
limit on the mass-spin relation for the central black hole in
microquasars~\citep{Ter-Abr-Klu:2005:ASTRA:QPOresmodel,Tor:2005:ASTRN:,Tor-etal:2006:ProcAEC:toapp}.

\citet{Asch:2004:ASTRA:,Asc:2006:CHIAA:,Asch:2007:CHIAA:MassSpin} discovered
that two changes of sign of the radial gradient of the Keplerian orbital
velocity as measured in the locally non-rotating
frame~\citep[LNRF,][]{Bar-Pre-Teu:1972:ASTRJ2:} occur in the equatorial plane
of Kerr black holes with $a>0.9953$. \citet{Stu-etal:2004:GRQC:} have found
that the gradient sign change in the LNRF-velocity profiles occurs also for
non-geodesic motion with uniform distribution of the specific angular momentum
$\ell(r,\theta)=\mathrm{const}$ (i.e., in marginally stable thick discs)
around extremely rapid Kerr black holes with $a>0.99979$.\footnote{Note that
  the assumption of uniform distribution of the specific angular momentum can
  be relevant at least at the inner parts of the thick disc and that matter in
  the disc follows nearly geodesic circular orbits nearby the center of the
  disc and in the vicinity of its inner edge determined by the cusp of its
  critical equipotential surface~\citep[see][]{Abr-Jar-Sik:1978:ASTRA:}.}  The
global character of the phenomenon is given in terms of topology changes of
the von Zeipel surfaces (equivalent to equivelocity surfaces in the tori with
$\ell (r,\theta)=\mathrm{const}$). Toroidal von Zeipel surfaces exist around
the circle corresponding to the minimum of the equatorial LNRF-velocity
profile, indicating possibility of development of some instabilities in that
part of the marginally stable disc with positive gradient of the orbital
velocity in
LNRF~\citep{Stu-Sla-Tor:2004:RAGtime4and5:CrossRef,Stu-etal:2004:GRQC:,Stu-Sla-Ter:2006:PoS:Humpy,Stu-Sla-Tor:2006:AECIC2005:CrossRef,Stu-Kot-Tor:2007:RAGtime8and9ThisVol:MrmQPO,Stu-Sla-Ter:2006:ASTRA:Humpy,Stu-Sla-Tor:2007:IAU26:CrossRef,Stu-Sla-Tor:2007:ASTRA:}.

The positive radial gradient of orbital LNRF-velocity around black holes with
$a>0.9953$ seems to be a physically interesting phenomenon, even if a direct
mechanism relating this phenomenon to triggering the oscillations, and
subsequent linking of the oscillations to the excitation of radial (and
vertical) epicyclic oscillations, is unknown. Therefore, an extended orbital
resonance model (EXORM) has been developed, with hypothetical hump-induced
oscillations assumed to enter a non-linear resonance with the radial or
vertical epicyclic oscillations~\citep{Stu-Sla-Ter:2006:ASTRA:Humpy}. It
should be stressed that due to the non-linear resonance, combinational
frequencies are allowed to be observable.

In the EXORM, the frequency of the hump-induced oscillations is related to
the maximal positive radial gradient of the LNRF-velocity in the ``humpy''
velocity profile in the general relativistic, coordinate-independent form.
Further, since the gradient is defined locally, being connected to the LNRF,
it has to be transformed into the form related to distant stationary
observers, giving observationally relevant ``humpy'' frequency
$\nu_\mathrm{h}$. Then the ``humpy'' and epicyclic frequencies could be
estimated at the radius of definition of the ``humpy'' frequency.

In the case of Keplerian discs, the epicyclic resonance radii $r_{3:1}$ and
$r_{4:1}$ (with $\nu_{\mathrm{v}}\!:\!\nu_{\mathrm{r}}=3\!:\!1$, $4\!:\!1$)
are located in vicinity of the ``humpy'' radius $r_{\mathrm{h}}$ where
efficient triggering of oscillations with
frequencies${}\!\sim\nu_{\mathrm{h}}$ could be expected.  Asymptotically (for
$1-a<10^{-4}$) the ratio of the epicyclic and Keplerian frequencies and the
humpy frequency is nearly constant, i.e., almost independent of $a$, being for
the radial epicyclic frequency $\nu_\mathrm{r}\!:\!\nu_\mathrm{h} \sim
3\!:\!2$.  In the case of thick discs, the situation is more complex due to
dependence on distribution of the specific angular momentum $\ell$ determining
the disc properties. For $1-a<10^{-6}$, the frequency ratios of the humpy
frequency and the orbital and epicyclic frequencies are again nearly constant
and independent of both $a$ and $\ell$ being for the radial epicyclic
frequency $\nu_\mathrm{r}\!:\!\nu_\mathrm{h}\sim 4\!:\!1$.  In the limiting
case of very slender tori ($\ell\sim\ell_{\mathrm{ms}}$) the epicyclic
resonance radius $r_{4:1}\sim r_{\mathrm{h}}$ for all the relevant interval of
$1-a<2\times 10^{-4}$.

In Section~\ref{LNRFvprof}, we briefly summarize properties of the Aschenbach
effect for Keplerian thin discs, and $\ell = \mathrm{const}$ thick discs. In
Section~\ref{xormodel}, the extended resonance model is introduced, i.e., the
critical ``humpy'' frequency, connected to the LNRF-velocity positive gradient
in the humpy profiles, is given in the physically relevant, coordinate
independent form for the both Keplerian and $\ell = \mathrm{const}$ discs. At
the radius of its definition, the humpy frequency is compared to the radial
and vertical epicyclic frequency and the orbital frequency. In
Section~\ref{xresmapp}, fitting of the observed frequencies in the
GRS~1915$+$105 microquasar in the framework of the EXORM is summarized, while
it is demonstrated that the data reported for Sgr\,A$^*$ could not be probably
fitted by EXORM. In Section \ref{conclus}, concluding remarks are presented.

\section{LNRF-velocity profiles of accretion discs}\label{LNRFvprof}

The locally non-rotating frames (LNRF) are given by the tetrad of
1-forms~\citep{Bar-Pre-Teu:1972:ASTRJ2:}
\begin{equation}
\begin{aligned}
  \vec{e}^{(t)} &=\left(\frac{\Sigma\Delta}{A}\right)^{1/2}\,\vec{d}t\,,&
    \vec{e}^{(\phi)} &=\left(\frac{A}{\Sigma}\right)^{1/2}\sin\theta
      (\vec{d}\phi-\omega\,\vec{d}t)\,,\\
  \vec{e}^{(r)} &=\left(\frac{\Sigma}{\Delta}\right)^{1/2}\,\vec{d}r\,,&
    \vec{e}^{(\theta)} &=\Sigma^{1/2}\,\vec{d}\theta\,,
\end{aligned}                                                   \label{e14}
\end{equation}
where
\begin{equation}
  \omega = -\frac{g_{t\phi}}{g_{\phi\phi}} = \frac{2ar}{A}      \label{e18}
\end{equation}
is the angular velocity of the LNRF relative to distant observers. 

In the Kerr spacetimes with the rotational parameter assumed to be $a>0$, the
relevant metric coefficients in the standard Boyer--Lindquist coordinates
read:
\begin{equation}
  g_{tt} = -\frac{\Delta - a^2 \sin^2 \theta}{\Sigma}\,,\:
  g_{t\phi} = -\frac{2ar\sin^2 \theta}{\Sigma}\,,\:
  g_{\phi\phi} = \frac{A\sin^2 \theta}{\Sigma}\,,\:
  g_{rr} = \frac{\Sigma}{\Delta}\,,\:
  g_{\theta\theta} = \Sigma\,,
\end{equation}
where
\begin{equation}
  \Delta = r^2-2r+a^2\,,\quad
  \Sigma = r^2+a^2 \cos^2 \theta\,,\quad
  A = (r^2+a^2)^2-\Delta a^2 \sin^2 \theta\,.
\end{equation}
The geometrical units, $c=G=1$, together with putting the mass of the black
hole $M=1$, are used in order to obtain completely dimensionless formulae
hereafter.

For matter orbiting a Kerr black hole with a 4-velocity $U^{\mu}$ and angular
velocity profile $\Omega(r,\theta)$, the azimuthal component of its 3-velocity
in the LNRF reads
\begin{equation}
  \vphi=\frac{U^{\mu} e^{(\phi)}_{\mu}}{U^{\nu} e^{(t)}_{\nu}} =
    \frac{A\sin\theta}{\Sigma\sqrt{\Delta}}\,
    (\Omega-\omega)\,.                                          \label{e19}
\end{equation}

\subsection{Keplerian thin discs}\label{Kepdisc}

In thin discs matter follows nearly circular geodetical orbits characterized
by the Keplerian distributions of the angular velocity and the specific
angular momentum (in the equatorial plane, $\theta=\pi/2$)
\begin{equation}
  \Omega=\Omega_{\mathrm{K}}(r;a)\equiv\frac{1}{(r^{3/2}+a)}\,,\qquad
  \ell=\ell_{\mathrm{K}}(r;a) \equiv
    \frac{r^2-2ar^{1/2}+a^2}{r^{3/2}-2r^{1/2}+a}\,.
\end{equation}
The azimuthal component of the Keplerian 3-velocity in the LNRF reads
\begin{equation}
  \vphi_{\mathrm{K}}(r;a)=\frac{(r^2+a^2)^2 - a^2\Delta
    - 2ar(r^{3/2}+a)}{r^2(r^{3/2}+a)\sqrt{\Delta}}              \label{e22}
\end{equation}
and formally diverges for $r\to r_{+}=1+\sqrt{1-a^2}$, where the \bh event
horizon is located. Its radial gradient is given by
\begin{equation}
  \pder{\vphi_{\mathrm{K}}}{r}
    =-\frac{r^5+a^4(3r+2)-2a^3r^{1/2}(3r+1)-2a^2 r^2(2r-5)+2ar^{5/2}(5r-9)}%
           {2\Delta^{3/2}\sqrt{r}\,(r^{3/2}+a)^2}\,.
\end{equation}
{\sloppy
  As shown by \citet{Asch:2004:ASTRA:,Asc:2006:CHIAA:}, the velocity profile
  has two changes of the gradient sign (where $\partial\vphi/ \partial r = 0$)
  in the field of rapidly rotating Kerr black holes with
  \mbox{$a>a_{\mathrm{c(K)}}\doteq 0.9953$}.
\par}

\subsection{Marginally stable tori}\label{margstab}

Perfect-fluid stationary and axisymmetric toroidal discs are characterized by
the 4-velocity field $U^{\mu} = (U^{t},0,0,U^{\phi})$ with
$U^{t}=U^{t}(r,\theta)$, $U^{\phi}=U^{\phi}(r,\theta)$, and by distribution of
the specific angular momentum $\ell=-U_{\phi}/U_{t}$.  The angular velocity of
orbiting matter, $\Omega=U^{\phi}/U^{t}$, is then related to $\ell$ by the
formula
\begin{equation}
  \Omega=-\frac{\ell g_{tt}+g_{t\phi}}{\ell g_{t\phi}+g_{\phi\phi}}\,.
\end{equation}

The marginally stable tori are characterized by uniform distribution of the
specific angular momentum
\begin{equation}
  \ell=\ell (r,\theta)=\mathrm{const}\,,
\end{equation}
and are fully determined by the spacetime structure through equipotential
surfaces of the potential $W=W(r,\theta)$ defined by the
relations~\citep{Abr-Jar-Sik:1978:ASTRA:}
\begin{equation}
  W-W_{\mathrm{in}}=\ln\frac{U_{t}}{(U_{t})_{\mathrm{in}}}\,,\qquad
  (U_t)^2=\frac{g_{t\phi}^2 - g_{tt}g_{\phi\phi}}{g_{tt}\ell^2
    +2g_{t\phi}\ell + g_{\phi\phi}}\,;                           \label{e6}
\end{equation}
the subscript ``in'' refers to the inner edge of the disc.  The LNRF orbital
velocity of the torus is given by
\begin{equation}
  \vphi_{\mathrm{T}}=\frac{A(\Delta-a^2\sin^2\theta) + 4a^2r^2\sin^2\theta}%
        {\Sigma\sqrt{\Delta}\,(A-2a\ell r)\sin\theta}\,\ell\,.  \label{e20}
\end{equation}

For marginally stable tori it is enough to consider the motion in the
equatorial plane, $\theta=\pi/2$. Formally, this velocity vanishes for
$r\to\infty$ and $r\to r_{+}$, i.e., there must be a change of its radial
gradient for any values of the parameters $a$ and $\ell$, contrary to the case
of Keplerian discs. The radial gradient of the equatorial LNRF velocity of
$\ell=\mathrm{const}$ tori reads
\begin{align}
  \pder{\vphi_{\mathrm{T}}}{r}
  =&\left\{
    \frac{[\Delta+(r-1)r][r(r^2+a^2)-2a(\ell-a)]}%
         {[r(r^2+a^2)-2a(\ell-a)]^2\sqrt{\Delta}}\right.\notag\\
   &-\left.\frac{r(3r^2+a^2)\Delta}%
   {[r(r^2+a^2)-2a(\ell-a)]^2\sqrt{\Delta}}\right\}\ell\,,      \label{e23}
\end{align}
so it changes its orientation at radii determined for a given $\ell$
by the condition
\begin{equation}                                               \label{e24}
     \ell=\ell_{\mathrm{ex}}(r;a) \equiv a +
     \frac{r^2[(r^2+a^2)(r-1)-2r\Delta]}{2a[\Delta+r(r-1)]}.
\end{equation}

\begin{figure}[p]
\begin{center}
\includegraphics[width=\linewidth,height=1.01\linewidth]{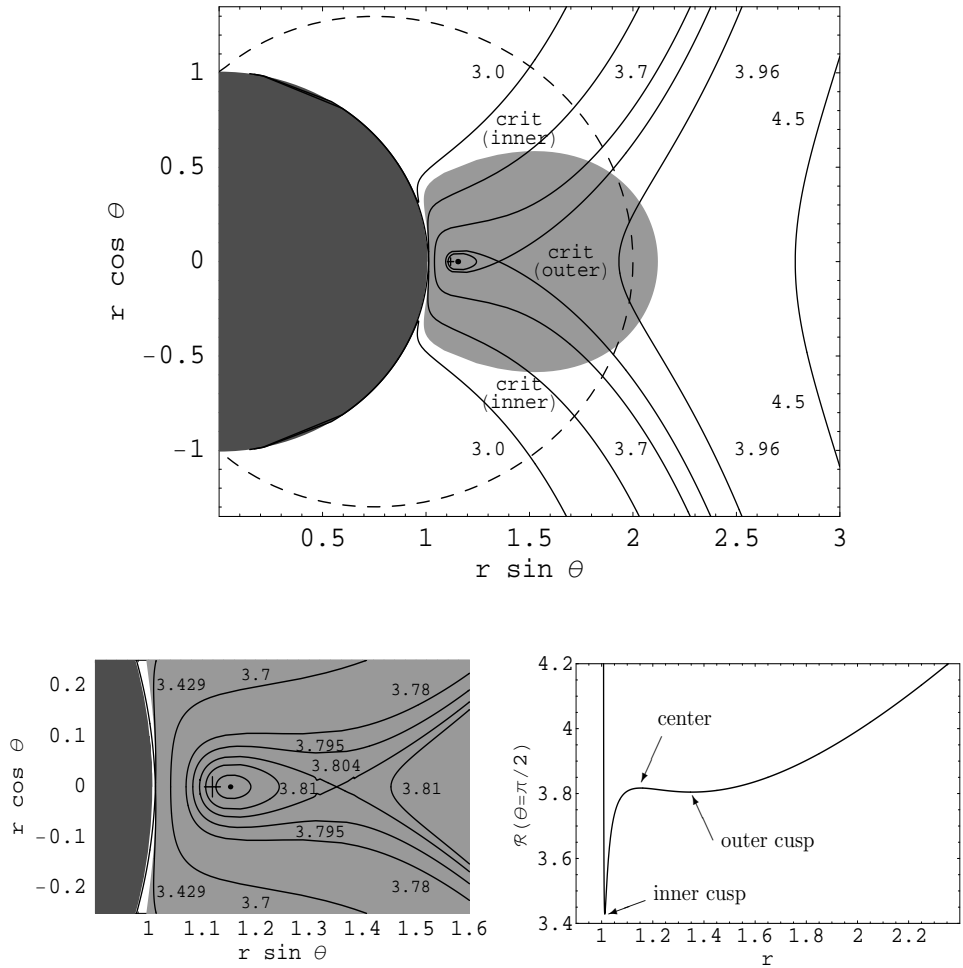}
\end{center}
\caption{\label{f2}Von Zeipel surfaces (meridional sections).
  For $a>a_{\mathrm{c(T)}}$ and $\ell$ appropriately chosen, two surfaces with
  a cusp, or one surface with both the cusps, together with closed (toroidal)
  surfaces, exist, being located always inside the ergosphere (dashed surface)
  of a given spacetime.  Both the outer cusp and the central ring of closed
  surfaces are located inside the toroidal equilibrium configurations
  corresponding to marginally stable thick discs (light-gray region; its shape
  is determined by the critical self-crossing \emph{equipotential surface} of
  the potential $W(r,\theta)$. The cross ($+$) denotes the centre of the
  torus. Dark region corresponds to the black hole. Figures illustrating all
  possible configurations of the von Zeipel surfaces are presented
  in~\citet{Stu-etal:2004:GRQC:}. Here we present the figure plotted for the
  parameters $a=0.99998$, $\ell=2.0065$. Critical value of the von Zeipel
  radius corresponding to the inner and the outer self-crossing surface is
  $\mathcal{R}_{\mathrm{c(in)}}\doteq 3.429$ and
  $\mathcal{R}_{\mathrm{c(out)}}\doteq 3.804$, respectively, the central ring
  of toroidal surfaces corresponds to the value
  $\mathcal{R}_{\mathrm{center}}\doteq 3.817$. Interesting region containing
  both the cusps and the toroidal surfaces is plotted in detail at the left
  lower figure. Right lower figure shows the behaviour of the von Zeipel
  radius in the equatorial plane. \citep[Taken
  from][]{Stu-Sla-Ter:2006:ASTRA:Humpy}}
\end{figure}

For both thick tori and Keplerian discs we have to consider the limit on the
disc extension given by the innermost stable orbit. For Keplerian discs this
is the marginally stable geodetical orbit, $r_{\mathrm{in}}\approx
r_{\mathrm{ms}}$, while for thick tori this is an unstable circular geodesic
kept stable by pressure gradients and located between the marginally bound and
the marginally stable geodetical orbits, $r_{\mathrm{mb}}\lesssim
r_{\mathrm{in}}\lesssim r_{\mathrm{ms}}$, with the radius being determined by
the specific angular momentum $\ell=\mathrm{const}\in
(l_{\mathrm{ms}},l_{\mathrm{mb}})$ through the equation $\ell=
\ell_{\mathrm{K}}(r;a)$; $\ell_{\mathrm{ms}}$ ($\ell_{\mathrm{mb}}$) denotes
specific angular momentum of the circular marginally stable (marginally bound)
geodesic.

Detailed discussion of~\citet{Stu-etal:2004:GRQC:} shows that two physically
relevant changes of sign of $\partial\vphi_{\mathrm{T}}/\partial r$ in the
tori occur for Kerr black holes with the rotational parameter
$a>a_{\mathrm{c(T)}}\doteq 0.99979$. The interval of relevant values of the
specific angular momentum $\ell\in
(\ell_{\mathrm{ms}}(a),\ell_{\mathrm{ex(max)}}(a))$, where
$\ell_{\mathrm{ex(max)}}(a)$ corresponds to the local maximum of the
function~(\ref{e24}), grows with $a$ growing up to the critical value of
$a_{\mathrm{c(mb)}}\doteq 0.99998$. For $a>a_{\mathrm{c(mb)}}$, the interval
of relevant values of $\ell\in (\ell_{\mathrm{ms}}(a),\ell_{\mathrm{mb}}(a))$
is narrowing with the rotational parameter growing up to $a=1$, which
corresponds to a singular case where
$\ell_{\mathrm{ms}}(a=1)=\ell_{\mathrm{mb}}(a=1)=2$.  Notice that the
situation becomes to be singular only in terms of the specific angular
momentum; it is shown \citep[see][]{Bar-Pre-Teu:1972:ASTRJ2:} that for $a=1$
both the total energy $E$ and the axial angular momentum $L$ differ at
$r_{\mathrm{ms}}$ and $r_{\mathrm{mb}}$, respectively, but their combination,
$\ell\equiv L/E$, giving the specific angular momentum, coincides at these
radii.

A physically reasonable global quantity characterizing rotating fluid
configurations in terms of the LNRF orbital velocity is so-called von Zeipel
radius defined by the relation
\begin{equation}
  \mathcal{R}\equiv\frac{\ell}{\vphi_{\mathrm{LNRF}}}
    =(1-\omega\ell)\,\tilde{\rho}\,,                            \label{e36}
\end{equation}
which generalizes in another way as compared
with~\citep{Abr-Nur-Wex:1995:CLAQG:} the Schwarz\-schild\-ian definition of the
gyration radius $\tilde{\rho}$~\citep{Abr-Mil-Stu:1993:PHYSR4:}.  Note that,
except for the Schwarzschild case $a=0$, the von Zeipel surfaces, defined as
the surfaces of $\mathcal{R}(r,\theta;a,\ell) = \mathrm{const}$, \emph{do not
  coincide} with those introduced by~\citet{Koz-Jar-Abr:1978:ASTRA:} as the
surfaces of constant $\ell/\Omega$~\citep[see][ for more
details]{Stu-etal:2004:GRQC:}.

In the case of marginally stable tori the von Zeipel surfaces
$\mathcal{R}=\mathrm{const}$ coincide with the equivelocity surfaces
$\vphi(r,\theta;a,\ell)= \vphi_{\mathrm{T}} = \mathrm{const}$.  Topology of
the von Zeipel surfaces can be directly determined by the behaviour of the von
Zeipel radius in the equatorial plane
\begin{equation}
  \mathcal{R}(r,\theta=\pi/2;a,\ell)
    =\frac{r(r^2+a^2)-2a(\ell-a)}{r\sqrt{\Delta}}\,.            \label{e39}
\end{equation}
The local minima of the function~(\ref{e39}) determine loci of the cusps of
the von Zeipel surfaces, while its local maximum (if it exists) determines a
circle around which closed toroidally shaped von Zeipel surfaces are
concentrated (see Fig.\,\ref{f2}). Notice that the inner cusp is always
physically irrelevant being located outside of the toroidal configuration of
perfect fluid. Behaviour of the von Zeipel surfaces nearby the centre and the
inner edge of the thick discs orbiting Kerr black holes with
$a>a_{\mathrm{c(T)}}\doteq 0.99979$, i.e., the existence of the von Zeipel
surface with toroidal topology, suggests possible generation of instabilities
in both the vertical and radial direction.

In terms of the redefined rotational parameter $(1-a)$, the ``humpy'' profile
of the LNRF orbital velocity of marginally stable thick discs occurs for discs
orbiting Kerr black holes with $1-a<1-a_{\mathrm{c(T)}}\doteq 2.1\times
10^{-4}$, which is more than one order lower than the value
$1-a_{\mathrm{c(K)}}\doteq 4.7\times 10^{-3}$ found
by~\citet{Asch:2004:ASTRA:} for the Keplerian thin discs. Moreover, in the
thick discs, the velocity difference $\Delta\vphi_{\mathrm{T}}$ is smaller but
comparable with those in the thin discs. In fact, for $a \to 1$, the velocity
difference in the thick discs $\Delta\vphi_{\mathrm{T}}\approx 0.02$, while
for the Keplerian discs it goes even up to $\Delta\vphi_{\mathrm{K}}\approx
0.07$~\citep{Stu-Sla-Tor:2007:ASTRA:}.

\section{Extended orbital resonance model}\label{xormodel}

The orbital resonance model assumes non-linear parametric and forced
resonances of oscillations with the orbital (Keplerian) and radial or vertical
epicyclic frequencies, or their combinations. Here, we extend this model by
introducing hypothetical additional oscillations, induced by the hump in the
LNRF-velocity profile of accretion discs of both Keplerian and toroidal
character, that are supposed to be in a non-linear resonance with orbital or
epicyclic oscillations.

In Kerr spacetimes, the frequencies of the radial and latitudinal (vertical)
epicyclic oscillations related to an equatorial Keplerian circular orbit at a
given $r$ are determined by the formulae \citep[see,
e.g.,][]{Ali-Gal:1981:GENRG2:}
\begin{gather}
  \nu^{2}_{\mathrm{r}}
    = \nu^{2}_{\mathrm{K}}(1-6r^{-1} + 8ar^{-3/2} - 3a^2r^{-2})\,,\\ 
  \nu^{2}_{\mathrm{v}}
    \equiv\nu^{2}_{\theta}
    =\nu^{2}_{\mathrm{K}}(1-4ar^{-3/2} + 3a^2r^{-2})\,,
\end{gather}
where the Keplerian frequency $\nu_{\mathrm{K}} = \Omega_{\mathrm{K}}/2\pi$. A
detailed analysis of properties of the epicyclic frequencies can be found
in~\citet{Tor-Stu:2005:RAGtime6and7:CrossRef,Ter-Stu:2005:ASTRA:}. The
epicyclic oscillations with the frequencies $\nu_{\mathrm{r}}$,
$\nu_\mathrm{v}$ can be related to both the thin Keplerian
discs~\citep{Abr-Klu:2000:ASTRAL:,Kat:2004:PUBASJ:mass} and thick, toroidal
discs~\citep{Rez-etal:2003:MONNR:}.

According to~\citet{Asch:2004:ASTRA:,Asc:2006:CHIAA:}, the non-monotonicity of
the LNRF-velocity profile of accretion discs could excite oscillations with
characteristic frequency that has to be related to the maximum gradient in the
``humpy'' part of the accretion discs velocity~profile.

Although there is no detailed idea on the mechanism generating the
hump-induced oscillations, it is clear that the Aschenbach proposal of
defining the characteristic frequency deserves attention.  It should be
stressed, however, that a~detailed analysis of the instability could reveal
a~difference between the characteristic frequency and the actual observable
one, as the latter should be associated with the fastest growing unstable
mode.  In any case, the humpy frequency represents an upper limit on the
frequencies of the hump-induced oscillations, as it is given by maximum of the
LNRF-velocity gradient in the humpy part of the velocity profile.

At the present state of the EXORM, we assume that the characteristic humpy
frequency is a typical frequency of oscillations induced by the conjectured
``humpy instability,'' and that the humpy oscillations could excite
oscillations with the epicyclic frequencies or some combinational frequencies,
if appropriate conditions for a~forced resonance are satisfied in vicinity of
the radius where the humpy oscillations occur~\citep{Stu-Sla-Tor:2007:ASTRA:}.

The fully general relativistic definition of the critical frequency for a
possible excitation of oscillations in the disc is given by the relations
\begin{equation}
  \nu^{\tilde{R}}_{\mathrm{crit}}
    = \left.\frac{\partial\vphi}{\partial\tilde{R}}\right|_{\mathrm{max}}\,,
  \qquad
  \dif\tilde{R}
    =\sqrt{g_{rr}}\,\dif r
    =\sqrt{\frac{\Sigma}{\Delta}}\,\dif r\,,                   \label{eq23}
\end{equation}
where $\vphi = \vphi_{\mathrm{K}}(r;a)$ in thin Keplerian discs, and $\vphi =
\vphi_{\mathrm{T}}(r;l,a)$ in marginally stable thick discs and $\tilde{R}$ is
the physically relevant, coordinate independent proper radial distance.  Such
a locally defined frequency, confined naturally to the observers orbiting the
black hole with the LNRF, should be further related to distant stationary
observers by the formula (taken at the BL coordinate $r$ corresponding to
$(\partial\vphi/\partial\tilde{R})_{\mathrm{max}})$
\begin{equation}
  \nu_{\mathrm{h}}=\nu^{\tilde{R}}_{\infty}
    = \sqrt{-(g_{tt}+2\omega g_{t\phi}+\omega^2 g_{\phi\phi})}\,
      \nu^{\tilde{R}}_{\mathrm{crit}}\,.                     \label{rce_24}
\end{equation}
We call such a coordinate-independent and, in principle, observable frequency
the ``humpy frequency,'' as it is related to the humpy profile of $\vphi$, and
denote it $\nu_{\mathrm{h}}$.  It should be stressed that the physically
relevant humpy frequency $\nu_{\mathrm{h}} = \nu^{\tilde{R}}_{\infty}$,
connected to observations by distant observers and exactly defined by
Eqs~(\ref{eq23}) and (\ref{rce_24}), represents an upper limit on
characteristic frequencies of oscillations induced by the hump of the
LNRF-velocity profile, and the realistic humpy frequencies, as observed by
distant observers, can be expected close~to but smaller than
$\nu^{\tilde{R}}_{\infty}$.  Further, we denote $r_{\mathrm{h}}$ the BL radius
of definition of the humpy oscillations frequency, where
$\partial\vphi/\partial\tilde{R} =
(\partial\vphi/\partial\tilde{R})_{\mathrm{max}}$.

\begin{figure}[t]
\begin{center}
\includegraphics[width=.82\linewidth]{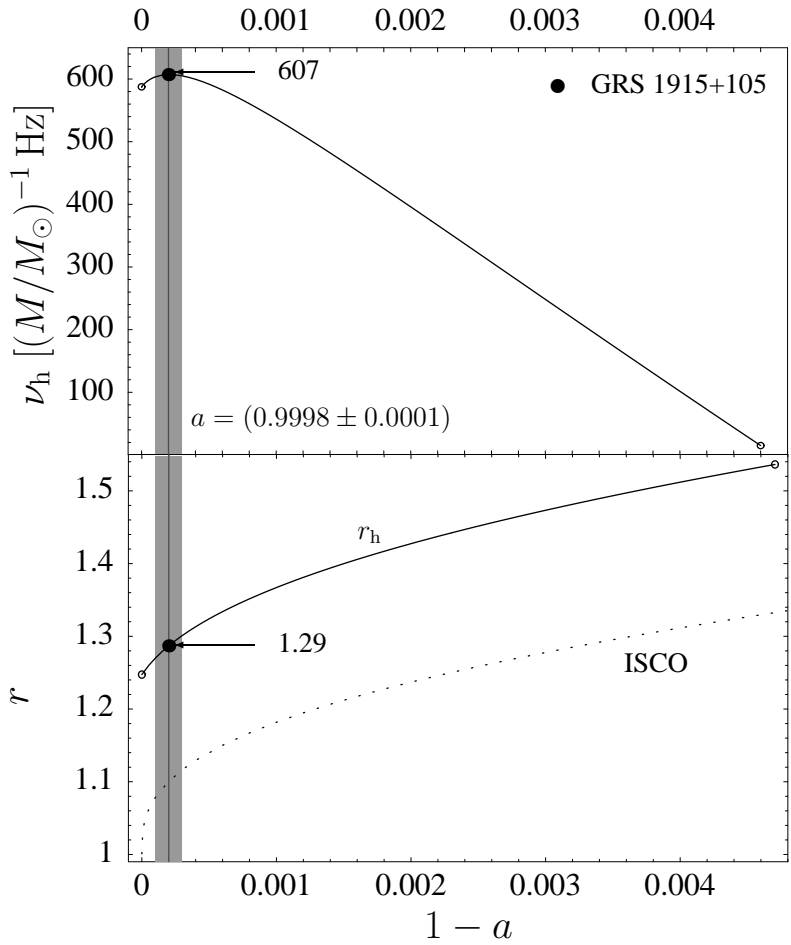}
\end{center}
\caption{\label{radfreq}Spin-dependence of the humpy frequency
  $\nu_{\mathrm{h}}$ and the humpy radius $r_{\mathrm{h}}$ that is compared
  with the Boyer--Lindquist radius of the innermost stable circular orbit.
  \citep[Taken from][]{Stu-Sla-Tor:2007:ASTRA:}}
\end{figure}

In the case of the Keplerian discs we obtain the ``humpy frequency'' to be
given by the relation
\begin{align}
  \nu_{\mathrm{h}}
    =&\frac{-r_{\mathrm{h}}^{5}-a^4(3r_{\mathrm{h}}+2)+2a^3
             r_{\mathrm{h}}^{1/2}(3r_{\mathrm{h}}+1)
            -2a^2 r_{\mathrm{h}}^{2} (2r_{\mathrm{h}}-5)
            +2ar_{\mathrm{h}}^{5/2}(5r_{\mathrm{h}}-9)}%
           {2\Delta_{\mathrm{h}}r_{\mathrm{h}}^{2}\,
            (r_{\mathrm{h}}^{3/2}+a)^2}\notag\\
    &\times\sqrt{r_{\mathrm{h}}-2
            -\frac{4a^2}{r_{\mathrm{h}}(r_{\mathrm{h}}^{2}+a^2)
            +2a^2}}\,,                                           \label{e1}
\end{align}
where $\Delta_{\mathrm{h}} = r_{\mathrm{h}}^2-2r_{\mathrm{h}}+a^2$.  The BL
radius $r_{\mathrm{h}}$ where the positive gradient of the velocity profile in
terms of the proper radial distance reaches its maximum, so-called ``humpy
radius,'' is given by the condition
\begin{equation}
  \pder{}{r}\left(\pder{\vphi}{\tilde{r}}\right )=0              \label{e2}
\end{equation}
leading to the equation
\begin{align}
3a^7(r+2)&+a^6\sqrt{r}(21r^2+18r-4)-a^5 r(33r^2+10r+20) \notag\\
  {}&+a^4 r\sqrt{r}(45r^3-62r^2-68r+16)-a^3 r^3(83r^2-122r-60) \notag\\
  {}&+a^2 r^4 \sqrt{r}(27r^2-130r+136)-9a r^5(7r^2-26r+24) \notag\\
  {}&+r^7 \sqrt{r}(3r-2)=0\,,                                    \label{e4}
\end{align}
which must be solved numerically. The spin dependence of the humpy radius and
the related humpy frequency is illustrated in Fig.\,\ref{radfreq}. The humpy
radius $r_{\mathrm{h}}$ falls monotonically with increasing spin $a$, while
the humpy frequency $\nu_{\mathrm{h}}$ has a maximum for $a=0.9998$, where
$\nu_{\mathrm{h(max)}}=607\,(\msun/M)\,\mathrm{Hz}$, and it tends to
$\nu_{{\mathrm{h}}({a\,\rightarrow1})}=588\,(\msun/M)\,\mathrm{Hz}$.

\begin{figure}[p]
\begin{center}
\includegraphics[width=.76\linewidth]{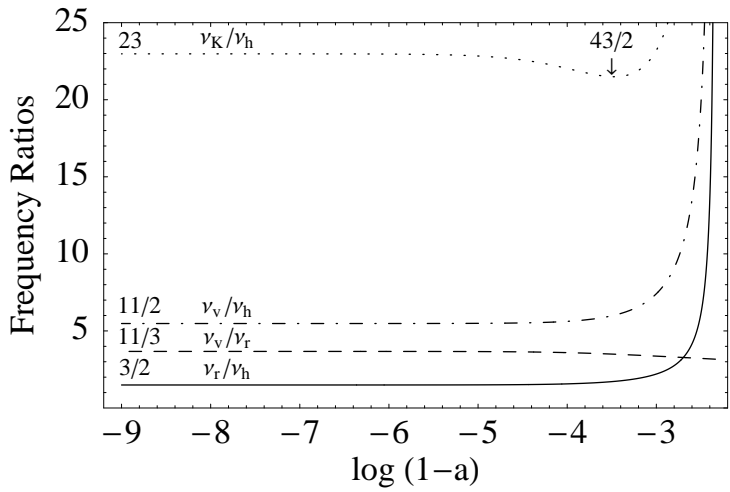}
\end{center}
\caption{\label{f8}Spin dependence of the ratios of the radial
  ($\nu_{\mathrm{r}}$) and vertical ($\nu_{\mathrm{v}}$) epicyclic
  frequencies, and the Keplerian frequency ($\nu_{\mathrm{K}}$) to the
  thin-disc humpy frequency related to distant observers ($\nu_{\mathrm{h}}$).
  All the frequency ratios are asymptotically (for $1-a<10^{-4}$) constant.
  There si $\nu_{\mathrm{K}}\!:\!\nu_{\mathrm{v}}\!:\!\nu_{\mathrm{r}}\!:\!
  \nu_{\mathrm{h}}\sim 46\!:\!11\!:\!3\!:\!2$. Therefore, we can expect some
  resonant phenomena on the ratio of
  $\nu_{\mathrm{r}}\!:\!\nu_{\mathrm{h}}\sim 3\!:\!2$, and
  $\nu_{\mathrm{K}}\!:\!\nu_{\mathrm{v}}\sim 4$ that could be both correlated.
  \citep[Taken from][]{Stu-Sla-Ter:2006:ASTRA:Humpy}}
\end{figure}

\begin{figure}[p]
\begin{center}
\includegraphics[width=.76\linewidth]{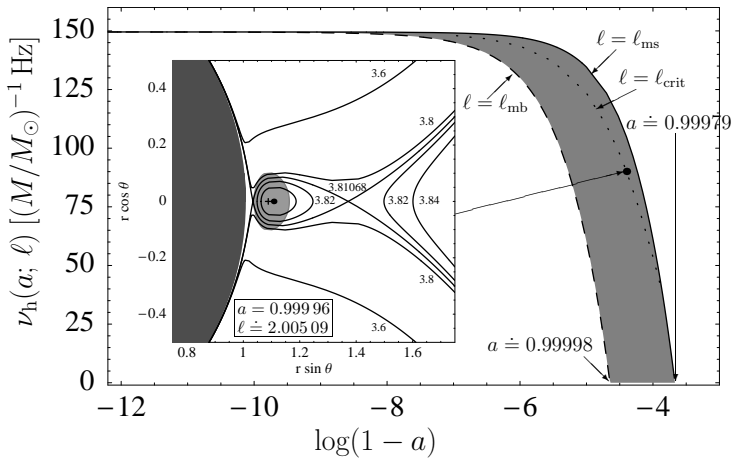}
\end{center}
\caption{\label{f9}Interval of humpy frequencies for the marginally stable
  thick discs with $\ell\in (\ell_{\mathrm{ms}},\ell_{\mathrm{mb}})$ as a
  function of the \bh spin $a$. For $a\to 1$, the interval is narrowing and
  asymptotically reaching the value of $150\,\mathrm{Hz}\,(M/\msun)^{-1}$.
  Dotted curve corresponds to the humpy frequencies of marginally stable
  slender tori with $\ell=\ell_{\mathrm{crit}}$, for which the critical
  von~Zeipel surface contains two cusps (as it is demonstrated for one special
  case in the left panel of the figure; the torus is given by the light-gray
  region).}
\end{figure}

\begin{figure}[t]
\begin{center}
\includegraphics[width=.77\linewidth]{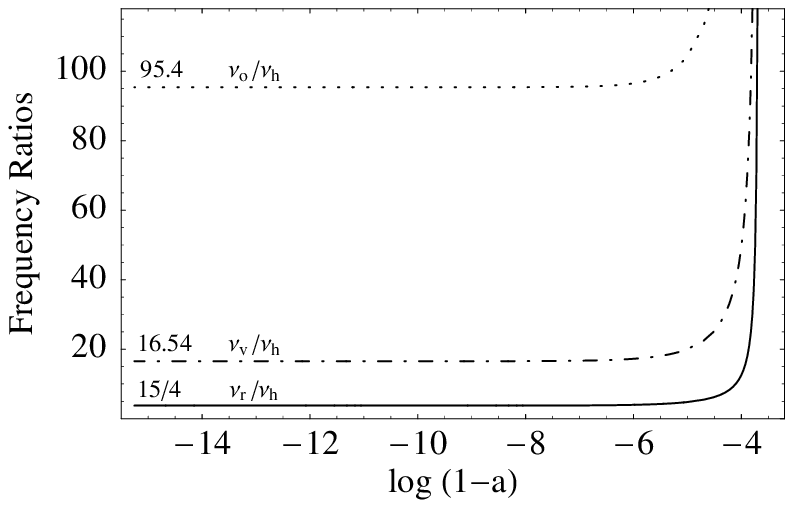}
\end{center}
\caption{\label{f10}Spin dependence of the ratios of the radial
  ($\nu_{\mathrm{r}}$) and vertical ($\nu_{\mathrm{v}}$) epicyclic
  frequencies, and the orbital frequency ($\nu_{\mathrm{o}}$) of the
  marginally stable $\ell=\ell_{\mathrm{ms}}$ disc to the thick-disc humpy
  frequency related to distant observers ($\nu_{\mathrm{h}}$).  All the
  frequency ratios are asymptotically (for $1-a<10^{-6}$) almost constant.
  \citep[Taken from][]{Stu-Sla-Ter:2006:ASTRA:Humpy}}
\end{figure}

The ratios of the humpy frequency and the orbital and epicyclic frequencies at
the humpy radius were determined in~\citet{Stu-Sla-Ter:2006:ASTRA:Humpy}
revealing almost spin-independent asymptotic behaviour for $a\to 1$
represented closely by the ratios of integer numbers,
$\nu_{\mathrm{K}}\!:\!\nu_{\mathrm{v}}\!:\!\nu_{\mathrm{r}}\!:\!
\nu_{\mathrm{h}}\sim 46\!:\!11\!:\!3\!:\!2$, which imply a possibility of
resonant phenomena between the hump-induced and orbital or epicyclic
oscillations. For Keplerian discs, the ratios of the epicyclic frequencies and
the humpy frequency are given in the dependence on the black-hole spin in
Fig.\,\ref{f8}.

The marginally stable tori have a structure that depends on the value of the
specific angular momentum $\ell\in(\ell_{\mathrm{ms}},\ell_{\mathrm{mb}})$.
The oscillations of slender tori ($\ell\approx\ell_{\mathrm{ms}}$) have
frequencies equal to the epicyclic frequencies relevant for test particle
motion, but the frequencies of non-slender tori are different, as shown for
pseudo-Newtonian tori~\citep{Sra:2005:ASTRN:,Bla-etal:2006:ASTRJ2:} and
expected for tori in the strong gravitational field of Kerr black holes.
Therefore, comparison of the humpy frequencies and the epicyclic frequencies
is relevant for the slender tori only.

The humpy frequency is defined for all $a>0.99979$ and all
$\ell\in(\ell_{\mathrm{ms}},\ell_{\mathrm{mb}})$, see Fig.\,\ref{f9}. It is
important that in the field of Kerr black holes with $1-a<10^{-8}$, there is
$\nu_{\mathrm{h}}(a,\ell)\simeq 150\,\mathrm{Hz}\,(M/\msun)^{-1}$
independently of $a$ and $\ell$~\citep{Stu-Sla-Ter:2006:ASTRA:Humpy}. Further,
the physically important case of tori admitting evolution of toroidal
von~Zeipel surfaces with the critical surface self-crossing in both the inner
and the outer cusps is allowed at $\ell=\ell_{\mathrm{crit}}$, where
$\ell_{\mathrm{crit}}\gtrsim\ell_{\mathrm{ms}}$ only slightly differs from
$\ell_{\mathrm{ms}}$, i.e., such tori can be slender, see Fig.\,\ref{f9}. The
ratios of $\nu_{\mathrm{r}}/\nu_{\mathrm{h}}$,
$\nu_{\mathrm{v}}/\nu_{\mathrm{h}}$ and $\nu_{\mathrm{o}}/\nu_{\mathrm{h}}$
are given for the tori with $\ell\approx\ell_{\mathrm{ms}}$ in
Fig.\,\ref{f10}.  Their asymptotical values, valid for $1-a<10^{-6}$, are
independent of both $a$ and $\ell$.

\begin{figure}[t]
\begin{minipage}{.48\linewidth}
\centering
\includegraphics[width=\linewidth]{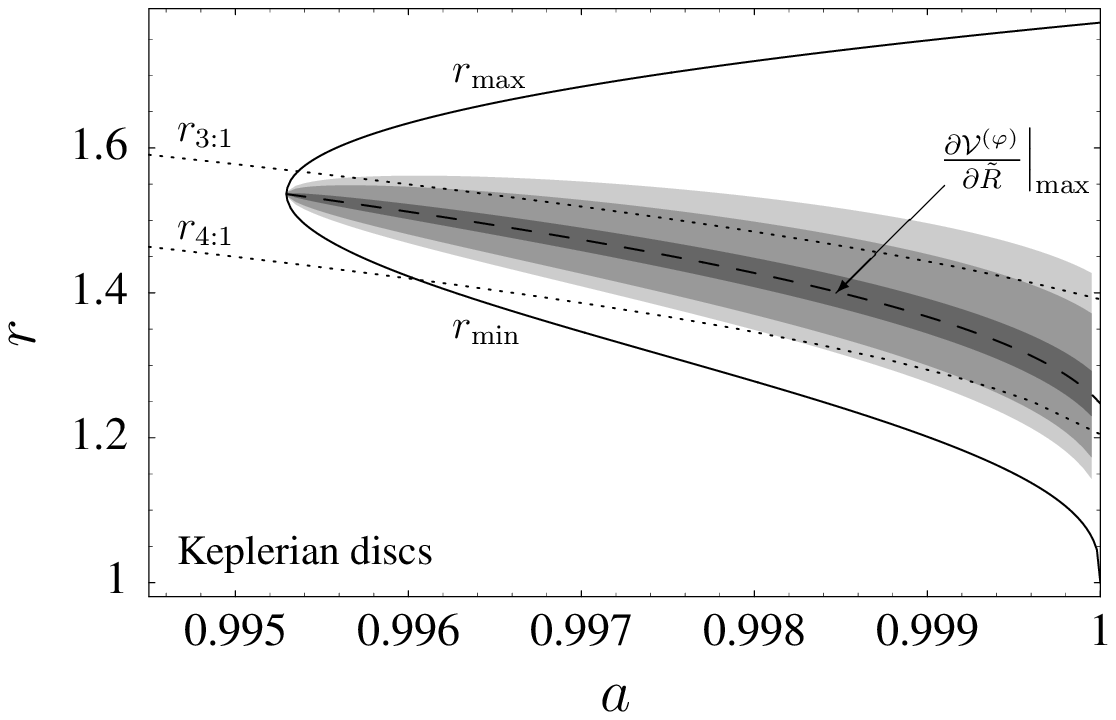}
\par\small (a)
\end{minipage}\hfill%
\begin{minipage}{.48\linewidth}
\centering
\includegraphics[width=\linewidth]{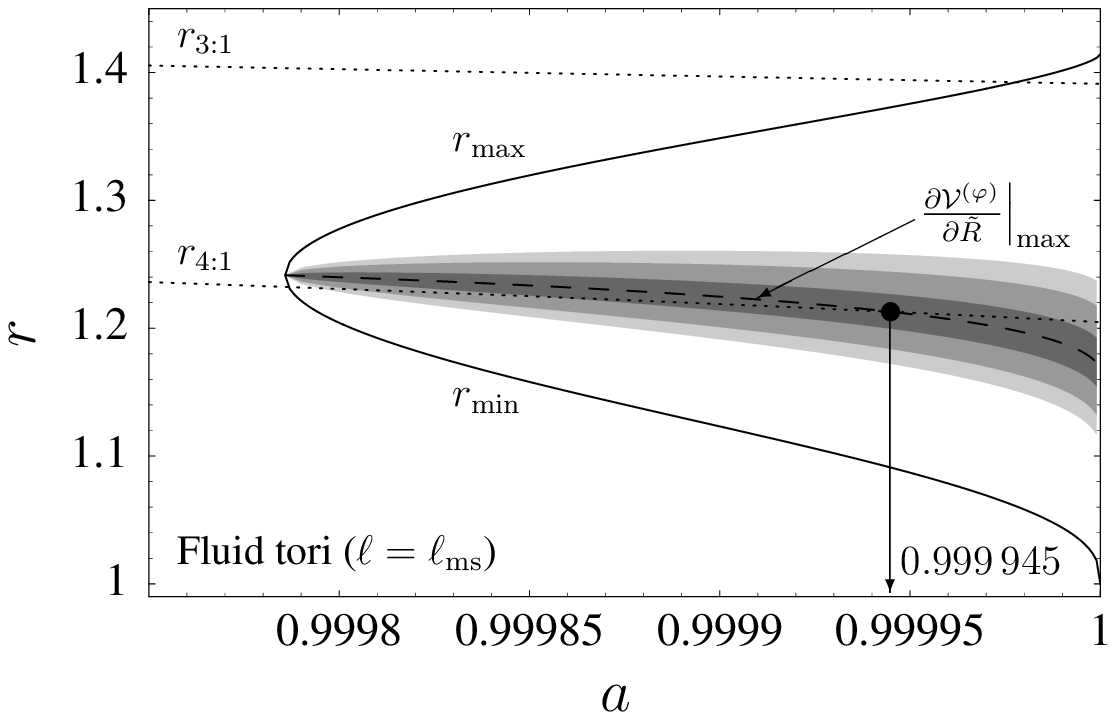}
\par\small (b)
\end{minipage}
\caption{\label{f4}Positions of local extrema of $\vphi$ (in BL
  coordinates) for Keplerian discs~(a) and marginally stable discs with
  $\ell=\ell_{\mathrm{ms}}$~(b) together with the locations of resonant orbits
  $r_{3:1}$ and $r_{4:1}$ (where the resonance between the vertical and radial
  epicyclic oscillations takes place) in dependence on the rotational
  parameter $a$ of the black hole. Dashed curve corresponds to the maximum
  positive values of the LNRF orbital velocity gradient in terms of the proper
  radial distance where the critical frequency
  $\nu^{\tilde{R}}_{\mathrm{crit}}$ is defined, boundaries of shaded regions
  correspond to orbits where the velocity gradient giving the characteristic
  frequency, $\partial\vphi/\partial\tilde{R}$, reaches (a)~$99\,\%$,
  $90\,\%$, $80\,\%$ and (b)~$99\,\%$, $95\,\%$, $90\,\%$ of its maximum.
  \citep[Taken from][]{Stu-Sla-Ter:2006:ASTRA:Humpy}}
\end{figure}

Of course, in realistic situations the hump-induced oscillation mechanism
could work at the vicinity of $r_{\mathrm{h}}$, with slightly different
frequencies; we should take into account that the shift of the radius, where
the mechanism works, shifts both the locally measured (LNRF) frequency
(Eq.\,(\ref{eq23})) and the frequency related to distant observers
(Eq.\,(\ref{rce_24})). The zones of radii, where the critical frequency
$\nu^{\tilde{R}}_{\mathrm{crit}}$ differs up to $1\,\%$, $10\,\%$ and $20\,\%$
of its maximal value (given by
$(\partial\vphi/\partial\tilde{R})_{\mathrm{max}}$) for thin (Keplerian) discs
or $1\,\%$, $5\,\%$ and $10\,\%$ of its maximum for marginally stable discs
with $\ell=\ell_{\mathrm{ms}}$, are given in Fig.\,\ref{f4}.  We can see
(Fig.\,\ref{f4}) that the resonant epicyclic frequencies radii $r_{3:1}$ and
$r_{4:1}$ are located within the zone of the hump-induced oscillation
mechanism in both thin discs and marginally stable tori.

In Keplerian discs the sign changes of the radial gradient of the orbital
velocity in LNRF occur nearby the $r=r_{3:1}$ orbit (with
$\nu_{\mathrm{v}}\!:\!\nu_{\mathrm{r}}=3:1$), while in the vicinity of the
$r=r_{3:2}$ orbit (with $\nu_{\mathrm{v}}\!:\!\nu_{\mathrm{r}}=3\!:\!2$),
$\partial\vphi/\partial r<0$ for all values of $a$ for both Keplerian discs
and marginally stable tori with all allowed values of $\ell$.  The parametric
resonance, which is the strongest one for the ratio of the epicyclic
frequencies $\nu_{\mathrm{v}}\!:\!\nu_{\mathrm{r}}=3\!:\!2$, can occur at the
$r=r_{3:2}$ orbit, while its effect is much smaller at the radius $r=r_{3:1}$,
as noticed by~\citet{Abr-etal:2003:PUBASJ:}.  Nevertheless, the forced
resonance may take place at the $r_{3:1}$ orbit.  Notice that the forced
resonance at $r=r_{3:1}$ can generally result in observed QPOs frequencies
with $3\!:\!2$ ratio due to the beat frequencies allowed for the forced
resonance as shown in~\citet{Abr-etal:2004:RAGtime4and5:CrossRef}.  But the
forced resonance at $r_{3:1}$ between the epicyclic frequencies, induced by
the humpy profile of $\vphi$, seems to be irrelevant in the case of
microquasars, since all observed frequencies lead to the values of the
rotational parameter $a<a_{\mathrm{c(K)}}$, as shown
by~\citet{Ter-Abr-Klu:2005:ASTRA:QPOresmodel}.

\section{Application of the extended resonance model}\label{xresmapp}

The extended orbital resonance model with hump-induced oscillations can be
applied only to black hole systems containing a near-extreme Kerr black hole
candidates. One of the most promising of such systems seems to be the
microquasar GRS~1915$+$105, where the extremely high spin $a \sim 1$ was
predicted by continuous spectra fitting method~\citep{McC-etal:2006:ASTRJ2:}.
Another promising candidate for the near-extreme Kerr black hole could be
considered the Galaxy Centre Sgr\,A$^*$. In fact, all the QPO frequencies
observed in GRS~1915$+$105 microquasar could be explained in the framework of
the EXORM~\citep{Stu-Sla-Tor:2007:ASTRA:,Sla-Stu:2007:RAGtime8and9ThisVol:EXORM}.
Therefore, we briefly summarize these results, and then we consider the case
of Sgr\,A$^*$, assuming relevance of all the three frequencies reported by
Aschenbach~\citep{Asch:2004:ASTRA:,Asch:2007:CHIAA:MassSpin}.

\subsection{GRS~1915$\boldsymbol{+}$105}\label{grs}

Assuming mass $M=14.8\,\msun$ and dimensionless spin $a=0.9998$ for the
GRS~1915$+$105 Kerr black hole~\citep[see][ for
details]{Stu-Sla-Tor:2007:ASTRA:,Sla-Stu:2007:RAGtime8and9ThisVol:EXORM}, the
EXORM predicts the following pattern of observable frequencies composed from
the humpy and epicyclic frequencies and their combinations
\begin{gather}
  \nu_{1}\sim(\nu_{\mathrm{r}}-\nu_{\mathrm{h}})
    =(26\pm 2)\,\mathrm{Hz}\,,                                  \label{e12}\\
  \nu_{\mathrm{h}}\equiv\nu_{2}
    =(41\pm 1)\,\mathrm{Hz}\,,                                   \label{e8}\\
  \nu_{\mathrm{r}}\equiv\nu_{3}
    =(67\pm 1) \,\mathrm{Hz}\,,                                  \label{e9}\\
  \nu_{4}\sim(\nu_{\mathrm{r}}+\nu_{\mathrm{h}})
    =(108\pm 2)\,\mathrm{Hz}\,,                                 \label{e10}\\
  \nu_{5}\sim(\nu_{\mathrm{v}}-\nu_{\mathrm{r}})
    =(0.17\pm 0.01)\,\mathrm{kHz}\,.                            \label{e11}
\end{gather}

The corresponding humpy radius is $r_{\mathrm{h}}= 1.29$.  At such a~radius,
the vertical epicyclic frequency of a~particle orbiting the Kerr black hole
with the mass and spin given above reaches the value
$\nu_{\mathrm{v}}=(0.23\pm 0.01)\,\mathrm{kHz}$ that enters the highest
(combinational) frequency.

\subsection{Sgr\,A$^{\boldsymbol{*}}$}\label{SgrA}

There are three frequencies related to the Sgr\,A$^*$
QPOs~\citep{Asch:2004:ASTRA:,Tor:2005:ASTRN:}:
\begin{equation}
  \nu_{\mathrm{u}}=1.445\,\mathrm{mHz}\,,\quad
  \nu_{\mathrm{m}}=0.885\,\mathrm{mHz}\,,\quad
  \nu_{\mathrm{l}}=0.459\,\mathrm{mHz}\,. 
\end{equation}
These frequencies come in the rational ratio
$\nu_{\mathrm{u}}\!:\!\nu_{\mathrm{m}}\!:\!\nu_{\mathrm{l}} \sim
3\!:\!2\!:\!1$.  Assuming EXORM with the humpy and radial epicyclic frequency
and their combinational frequencies, we can distinguish three different cases
of the resonant phenomena explaining the observed frequencies.

\subsubsection*{$\boldsymbol{\nu_{\mathrm{r}}\!:\!\nu_{\mathrm{h}}
  \sim 3\!:\!2}$}

The observed frequency pattern is given by
$\nu_{\mathrm{u}}=\nu_{\mathrm{r}}$, $\nu_{\mathrm{m}}=\nu_{\mathrm{h}}$,
$\nu_{\mathrm{l}}=\nu_{\mathrm{r}}-\nu_{\mathrm{h}}$.  The value of the spin,
given by the frequency ratio, and the humpy frequency are then given by
(compare Figs~\ref{radfreq} and~\ref{f8})
\begin{equation}
  a_{3:2}=0.999984\,,\qquad
  \nu_{\mathrm{h(3:2)}} = 590\,\frac{\msun}{M}\,\mathrm{Hz}\,.
\end{equation}
Using the condition $\nu_{\mathrm{m}}=\nu_{\mathrm{h}}$, we obtain mass of the
black hole to be
\begin{equation}
  M=0.667\times10^{6}\,\msun\,.
\end{equation}
(Note that in this case we have chosen~-- see Fig.\,\ref{f8}~-- the value of
the spin at the region where the frequency ratio $\nu_{\mathrm{r}}\!:\!
\nu_{\mathrm{h}}$ starts to be close to the asymptotical value of${}\sim
3\!:\!2$.)

\subsubsection*{$\boldsymbol{\nu_{\mathrm{r}}\!:\!\nu_{\mathrm{h}}
  \sim 2\!:\!1}$}

The observed frequency pattern is given by
$\nu_{\mathrm{u}}=\nu_{\mathrm{r}}+\nu_{\mathrm{h}}$,
$\nu_{\mathrm{m}}=\nu_{\mathrm{r}}$, $\nu_{\mathrm{l}}=\nu_{\mathrm{h}}$. The
value of the spin and the humpy frequency are then given by (compare
Figs~\ref{radfreq} and~\ref{f8})
\begin{equation}
  a_{2:1}=0.99925\,,\qquad
  \nu_{\mathrm{h(2:1)}} = 567\,\frac{\msun}{M}\,\mathrm{Hz}\,.
\end{equation}
Using the condition $\nu_{\mathrm{l}}=\nu_{\mathrm{h}}$, we obtain mass of the
black hole to be
\begin{equation}
  M=1.235\times10^{6}\,\msun\,.
\end{equation}

\subsubsection*{$\boldsymbol{\nu_{\mathrm{r}}\!:\!\nu_{\mathrm{h}}
  \sim 3\!:\!1}$}

The observed frequency pattern is given by $\nu_{\mathrm{u}}=\nu_{\mathrm{r}},
\nu_{\mathrm{m}}=\nu_{\mathrm{r}}-\nu_{\mathrm{h}},
\nu_{\mathrm{l}}=\nu_{\mathrm{h}}$. The value of the spin and the humpy
frequency are given by (Figs~\ref{radfreq} and~\ref{f8})
\begin{equation}
  a_{3:1}=0.999984\,,\qquad
  \nu_{\mathrm{h(3:1)}} = 425\,\frac{\msun}{M}\,\mathrm{Hz}\,.
\end{equation}
Using the condition $\nu_{\mathrm{l}}=\nu_{\mathrm{h}}$, we obtain mass of the
black hole to be
\begin{equation}
  M=0.926\times10^{6}\,\msun\,.
\end{equation}
However, by analysing data of the orbits of stars moving within 1000~light
hours of the Sgr\,A$^*$ black hole, its mass is estimated to be $M
\sim3.6\times10^{6}\,\msun$, and the error of the estimate is given
by~\citep{Ghe-etal:2005:ASTRJ2:,Wei-Mil-Ghe:2005:ASP338:astro-ph0512621}
\begin{equation}
  2.8\times10^{6}\,\msun<M<4.6\times10^{6}\,\msun.
\end{equation}
Clearly, the Sgr\,A$^*$ black hole mass predicted by the three simple variants
of the EXORM are all completely out of the mass estimates given by relatively
precise star-orbit measurements. Therefore, we can conclude that the EXORM
cannot be applied to explain the QPOs observed in the Sgr\,A$^*$ black hole
candidate.  Although we expect a fast rotating black hole in the Galaxy
Centre, it is probably not a near-extreme Kerr black hole that sites in
Sgr\,A$^*$.

\section{Concluding remarks}\label{conclus}

The extended orbital resonance model with the hypothetical humpy oscillations
could be related to the QPO resonant phenomena in both thin Keplerian discs
and marginally stable tori orbiting near-extreme Kerr black holes. The
non-linear resonance is assumed between oscillations with the humpy frequency
and the radial or vertical epicyclic frequency. Both parametric and forced
resonance phenomena are possible, therefore, the combinational frequencies are
allowed too.  Generally, more than two observable oscillations are predicted.

The EXORM can successfully explain all five QPO frequencies observed in the
microquasar GRS~1915$+$105~\citep{Stu-Sla-Tor:2007:ASTRA:}, where a
near-extreme black hole with $a \sim 1$ is predicted by spectral X-ray
continuum fitting~\citep{McC-etal:2006:ASTRJ2:}.
Although~\citet{Mid-etal:2006:MONNR:} refer to a substantially lower,
intermediate value of the black-hole spin, $a\sim0.7$, to which the EXORM
cannot be applied, the recent analysis
by~\citet{Nar-McC-Sha:2007:AstroCompObj:0710.4073} and
\citet{McC-Nar-Sha:2007:BH:0707.4492} demonstrated convincingly that the
near-extreme black hole is more probable there, making the predictions of the
EXORM still well viable. This model is successfully applied to explain QPOs
observed also in the case of the other microquasar XTE~J1650$-$500, and in the
near-extreme intermediate-mass Kerr black hole candidate in the system
NGC~5408~X-1~\citep{Sla-Stu:2007:RAGtime8and9ThisVol:EXORM}.

On the other hand, the EXORM is not able to explain, in any of its variants,
the QPOs observed in the Sgr\,A$^*$ source, i.e., in the Galactic Centre Kerr
black hole.  The differences in the black hole mass estimated by the EXORM and
by analysing the star orbits in vicinity of the black hole are too high to
believe that some more exact QPOs measurements could give a solution of this
discrepancy. On the other hand, it is worth to note that the Sgr\,A$^*$ QPOs,
if true, could be explained within the multiresonance model based on the
assumption of strong resonances of oscillation with the orbital and both
epicyclic frequencies~\citep{Stu-Kot-Tor:2007:RAGtime8and9ThisVol:MrmQPO}.
The spin estimated in this case is $\sim 0.983$, corresponding to a fast
rotating, but not near-extreme Kerr black hole.

We can conclude that the EXORM could be considered as a promising model of
QPOs in near-extreme Kerr black holes systems, where the oscillations occur in
the innermost parts of the accretion disc. The model enables very precise
measurements of the black hole parameters, in particular, of the black hole
spin since its value is given (usually) by the frequency ratio of the humpy
frequency and the radial epicyclic frequency. However, the predictions of the
EXORM have to be confronted to the black hole parameter estimates coming from
other methods, as those based on the optical phenomena in strong gravitational
fields, e.g., the spectral X-ray continuum fitting, the profiled spectral
(Fe-K) lines fitting, time delay methods, or by methods based on the orbital
motion analysis. The present situation seems to be rather controversial,
however, we believe that in future, with developing both the theoretical
models and observational techniques, we could be able to understand the
accretion phenomena in much deeper detail as compared to the present
understanding.

\ack

This work was supported by the \msmgrant{} and by the GA\v{C}R grant
202/06/0041 (Z.\,S.).

\providecommand{\uv}[1]{\glqq#1\grqq}
  \makeatletter\providecommand{\BibTeX}{B\hbox{\check@mathfonts
  \fontsize\sf@size\z@ \math@fontsfalse\selectfont
  IB}\TeX}\makeatother\providecommand{\cheatsort}[1]{} \let\origemph\emph
  \renewcommand{\emph}[1]{\origemph{\ #1}} \let\u\v

\end{document}